\documentclass[pra,showpacs,floatfix,twocolumn]{revtex4}
\usepackage{graphics}
\usepackage{epsfig}
\usepackage{amssymb}
\usepackage{color}
\usepackage{amsthm,amscd,amsfonts,amssymb,epsfig,graphicx}

\begin{document}

\date{\today }
\title{Dragging two-dimensional discrete solitons by moving linear defects}
\author{Valeriy A. Brazhnyi}
\email{brazhnyy@gmail.com}
\affiliation{Centro de F\'{\i}sica do Porto, Faculdade de Ci\^encias, Universidade do
Porto, R. Campo Alegre 687, Porto 4169-007, Portugal}
\author{Boris A. Malomed}
\email{malomed@post.tau.ac.il}
\affiliation{Department of Physical Electronics, School of Electrical Engineering,
Faculty of Engineering, Tel Aviv University, Tel Aviv 69978, Israel;\\
ICFO-Institut de Ciencies Fotoniques, and Universitat Politecnica de
Catalunya, Mediterranean Technology Park, 08860 Castelldefels (Barcelona),
Spain\thanks{%
temporary Sabbatical address}}

\begin{abstract}
We study the mobility of small-amplitude solitons attached to moving
defects which drag the solitons across a two-dimensional (2D)
discrete nonlinear-Schr\"{o}dinger (DNLS) lattice. Findings are
compared to the situation when a free small-amplitude 2D discrete
soliton is kicked in the uniform lattice. In agreement with
previously known results, after a period of transient motion the
free soliton transforms into a localized mode pinned by the
Peierls-Nabarro potential, irrespective of the initial velocity.
However, the soliton attached to the moving defect can be dragged
over an indefinitely long distance (including routes with abrupt
turns and circular trajectories) virtually without losses, provided
that the dragging velocity is smaller than a certain critical value.
Collisions between solitons dragged by two defects in opposite
directions are studied too. If the velocity is small enough, the
collision leads to a spontaneous symmetry breaking, featuring fusion
of two solitons into a single one, which remains attached to either
of the two defects.
\end{abstract}

\maketitle

\section{Introduction}

The discrete nonlinear Schr\"{o}dinger (DNLS)\ equations constitute a vast
class of systems which are profoundly interesting in their own right, and
serve as important physical models for nonlinear optics \cite{PhysRep},
matter waves in Bose-Einstein condensates (BECs) \cite{Smerzi}, and in other
contexts \cite{Panos}. In particular, soliton solutions to the DNLS equation
in one, two, and three dimensions (1D, 2D, and 3D) represent fundamental
localized modes in discrete media. Experimentally, 1D and 2D quasi-discrete
solitons have been created in nonlinear optical systems of several types,
see a comprehensive review in Ref. \cite{PhysRep}.

Local defects are important ingredients of DNLS models. They are interesting
as additional dynamical elements of the lattices \cite{Panos}, and have
direct physical realizations. In particular, they may describe various
strongly localized structures in photonic crystals \cite{defect0}, such as
nanocavities \cite{defect1}, micro-resonators \cite{defect2}, and quantum
dots \cite{defect3}. In the context of BEC, local defects can be easily
created and used for manipulating the condensate by means of focused laser
beams. This technique has made it possible to create optical tweezers for
trapping and controllable transfer of condensates by \emph{moving} optical
traps \cite{tweezers}. A scheme for nonlinear optical tweezers that can
extract solitons from a linear reservoir was proposed too \cite{Humberto}.

In a more general context, controllable transport of solitons in periodic
and quasiperiodic media, including the ultimate case of discrete lattices,
plays an important role in various applications, see Refs.~\cite{transport}
and references therein. Many attempts have been made to devise a simple way
of transferring solitons from one position to another with minimal losses.
In particular, dragging gap solitons in BEC, embedded into an optical
lattice, by a moving defect to which the soliton is attached, was proposed
as a means of the transport in Ref.~\cite{BBK}. Recently, manipulations of a
BEC vortex by a localized impurity representing a focused laser beam was
considered in Ref. \cite{Panos1}.

The subject of the present work is the dragging of discrete solitons in 2D
lattices by an attractive defect which plays the role of the mover. The
defect is shaped as a Gaussian of a finite width. After introducing the
model and recapitulating some relevant results for static trapped modes in
Sect. 2, we consider moving solitons in Sect. 3. First, the problem of the
immobility of free 2D DNLS solitons is briefly revisited, and then new
results are reported for the transfer of solitons by the moving attractive
defect. Both simple rectilinear dragging routes, and more sophisticated
ones, in the form of square, rhombic, and circular closed trajectories, are
considered. It is concluded that the trapped solitons may survive the
dragging over an indefinitely long distance, if the dragging velocity does
not exceed a certain critical value. The paper is concluded by Sect. 4.

\section{The model and stationary solutions}

\subsection{The formulation}

We consider the following model based on the 2D DNLS equation with a local
defect:
\begin{equation}
i\dot{u}_{n,m}+J\Delta _{2}u_{n,m}+V_{n,m}(t)u_{n,m}+\sigma
|u_{n,m}|^{2}u_{n,m}=0,  \label{timeDNLS}
\end{equation}%
where the overdot stands for the time derivative, $\Delta _{2}u_{n,m}\equiv
u_{n,m+1}+u_{n,m-1}+u_{n+1,m}+u_{n-1,m}-4u_{n,m}$ is the 2D discrete
Laplacian, the coupling constant of the lattice will be fixed by scaling, $%
J\equiv 1$, and $\sigma >0$ is the coefficient of the on-site
self-attraction. Discrete function $V_{n,m}$ account for the linear defect,
taken in the form as a Gaussians profile,
\begin{equation}
V_{n,m}(t)=v\exp \left\{ -\left[ \left( n-n(t)\right) ^{2}+\left(
m-m(t)\right) ^{2}\right] /\Delta ^{2}\right\} ,  \label{defect}
\end{equation}%
with respective strength $v$, width $\Delta $, and coordinates of the
(moving) center, $n(t)$ and $m(t)$. In this notation, positive and negative
strengths correspond to the attractive and repulsive defect, respectively.

Looking for stationary solutions in the case of the quiescent defect, with $%
m(t)=n(t)\equiv 0$ in Eq. (\ref{defect}),
\begin{equation}
u_{n,m}=U_{n,m}\exp (-i\omega t),  \label{omega}
\end{equation}%
we arrive at the nonlinear eigenvalue problem,
\begin{equation}
\omega U_{n,m}+\Delta _{2}U_{n,m}+V_{n,m}U_{n,m}+\sigma
|U_{n,m}|^{2}U_{n,m}=0,  \label{steadyDNLS}
\end{equation}%
for real frequency $\omega $ and the profile of the stationary discrete
mode, $U_{m,n}$, which may be complex, in the general case. In the absence
of the defect, the linearized version of Eq. (\ref{steadyDNLS}) gives rise
to the dispersion relation for linear modes $U_{m,n}\sim \exp \left(
ikm+iqn\right) $,%
\begin{equation}
\omega =4\left[ \sin ^{2}\left( k/2\right) +\sin ^{2}\left( q/2\right) %
\right] ,  \label{disp}
\end{equation}%
which features the phonon band, $0\leq \omega \leq 8$. Above and below the
band, full nonlinear equation (\ref{steadyDNLS}) gives rise to nonlinear
modes, described by respective curves $N(\omega )$, with $%
N=\sum_{m,n}|u_{m,n}|^{2}$ being the norm (power) of the mode. The important
difference between 2D and 1D settings is that, in the latter case, the
fundamental single-peak mode (alias the Sievers-Takeno discrete soliton)
admits the limit of $N\rightarrow 0$, while all the 2D solitons are bounded
by a critical norm, $N_{\mathrm{cr}}$, below which they do not exist \cite%
{Panos}. Accordingly, an abrupt delocalization (decay) of discrete
2D solitons was predicted in the case when the inter-site coupling
constant exceeds a certain critical value \cite{Bishop}. It is
worthy to mention that, introducing a defect with the linear and
nonlinear components and varying their strengths, or the lattice
coupling constant, in time (in the spirit of the ``soliton
management" techniques \cite{book}), one can change the critical
norm, $N_{\mathrm{cr}}$, and thus control the transition to the
delocalization \cite{we}.

\subsection{Existence curves for stationary solitons and stability}

\begin{figure}[th]
\epsfig{file=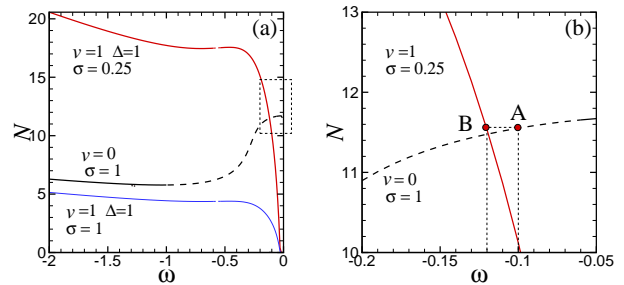,width=8cm} \caption{(Color online) (a) The
existence curves for on-site symmetric solitons in the defect-free
lattice ($v=0$, the black line), and in the
presence of the quiescent defect (\protect\ref{defect}) with $v=1$ and $%
\Delta =1$, for different values of the nonlinearity coefficient: $\protect%
\sigma =1$ and $\protect\sigma =0.25$ (the blue and red lines,
respectively). The dashed part of the black curve corresponds to unstable
solitons in the uniform lattice. Panel (b) shows a zoom of the dashed box
from (a). Points A and B correspond to the solitons with frequencies $%
\protect\omega _{A}=-0.1$ and $\protect\omega _{B}=-0.17$, and equal norms, $%
N_{A}=N_{B}$. }
\label{fig1}
\end{figure}

Curves $N(\omega )$ for the discrete solitons obtained from numerical
solutions of Eq. (\ref{steadyDNLS}), based on the continuation from the
anti-continuum limit [the one corresponding to $J=0$ in Eq. (\ref{timeDNLS}%
)], using the standard Newton's iteration routine, are displayed in Fig. \ref%
{fig1}, in the presence and in the absence of the defect. In panel \ref{fig1}%
(a) we compare three curves: the black one for the case of $\sigma =1$
without the defect, and the blue and red curves for $\sigma =1$ and $\sigma
=0.25$ with the defect of amplitude $v=1$ and width $\Delta =1$ [the red
curve is obtained from the blue one by rescaling of the norm, in order to
get a curve crossing the black one in a vicinity of $\omega \rightarrow 0$,
see Fig. \ref{fig1}(b)]. In the defect-free lattice, a family of stable
solitons (in agreement with the Vakhitov-Kolokolov (VK) criterion, it obeys
condition $dN/d\omega <0$ \cite{Panos}) can be found in the region of $%
-\infty <\omega <\omega _{\mathrm{cr}}\approx -1$. In this region, the
discrete soliton is tightly localized, being immobile, due to the strong
pinning to the underlying Peierls-Nabarro potential \cite{Panos}. The
Peierls-Nabarro potential becomes weak for broad small-amplitude discrete
solitons corresponding to $\omega \rightarrow 0$. Note that the introduction
of the positive defect can significantly reduce or completely suppress the
instability region, as shown by the blue and red curves in Fig. \ref{fig1}%
(a) where only a small segment of the existence curve in the interval of $%
-0.58<\omega <-0.52$ remains unstable. Note also that the attractive defect
removes the lower existence bound for the 2D discrete solitons, $N_{\mathrm{%
cr}}$, as the soliton goes over into a linear defect mode in the limit of $%
N\rightarrow 0$.

\begin{figure}[th]
\epsfig{file=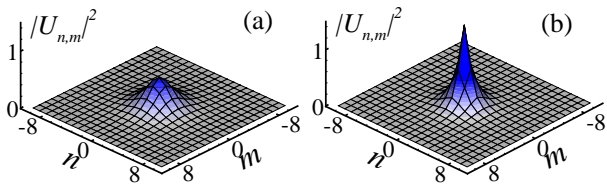,width=8cm}
\caption{Profiles of unstable and stable discrete solitons, which
correspond, respectively, to points A and B of the existence curves in Fig.
\protect\ref{fig1}(b).}
\label{fig2}
\end{figure}

According to the previous analysis, at small $|\omega |$ initially unstable
discrete solitons in the uniform lattice may be stabilized by switching the
attractive defect on \cite{Panos,we}. To keep the norm of the soliton
constant in this case, simultaneously with introducing the defect we
decrease the strength of the nonlinearity. As a result, we obtain the 
existence curve shown in red in Fig. \ref{fig1}(a), which features intersection with the
original black line in the region of $|\omega |\ll 1$. The stabilization
corresponds to the transition from point A to B in Fig. \ref{fig1}(b), which
follows the increase of the strength of the defect from $v=0$ to $v=1$ and
reduction of the strength of the nonlinear coefficient from $\sigma =1$ to $%
\sigma =0.25$. We have checked numerically that the transition between these
two configurations could be performed even instantaneously in time, leading
to a very small radiation loss. Examples of the profiles of the discrete
solitons found without and with the defect are displayed in Fig. \ref{fig2}.
Both solutions have the same norm, while solitons A and B belong,
respectively, to the unstable and stable branches of the existence curves,
cf. Fig. \ref{fig1}.

\section{Moving solitons}

\subsection{Dynamics of discrete solitons in the uniform lattice}

\begin{figure}[th]
\epsfig{file=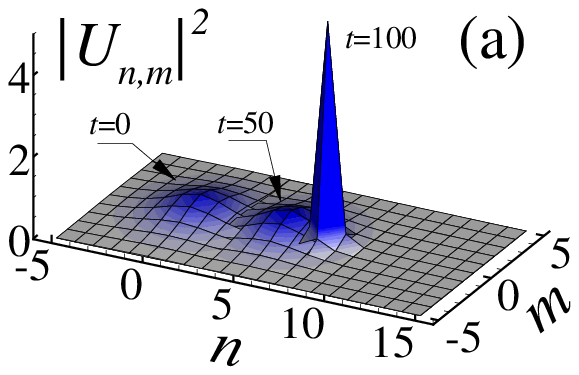,width=5cm} \epsfig{file=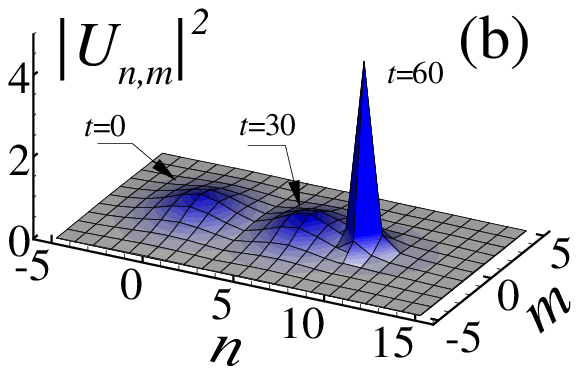,width=5cm} %
\epsfig{file=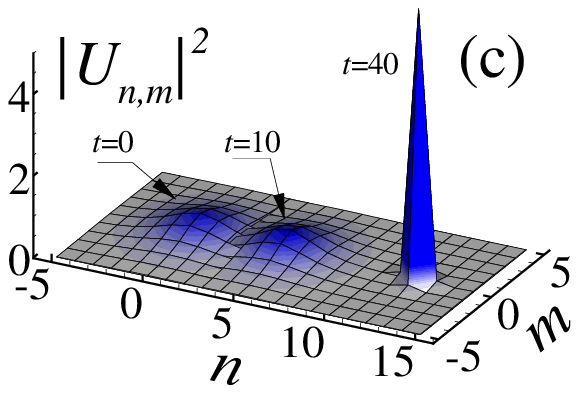,width=5cm}
\caption{Snapshots of the discrete soliton initially kicked in the
horizontal direction ($c_{m}=0$), are shown at different times. The initial
profile is taken from Fig. \protect\ref{fig2}(a). Parameters of the kick and
the coordinate of the final pinned site, $n_{f}$, are $c_{n}=0.1$, $n_{f}=7$
in (a); $c_{n}=0.2$, $n_{f}=9$ in (b), and $c_{n}=0.5$, $n_{f}=12$ in (c).}
\label{fig3}
\end{figure}

Proceeding to the consideration of traveling solitons, it is first relevant
to recapitulate basic results concerning the ballistic motion of kicked
solitons in the uniform lattice. We kick the discrete soliton by taking the
initial condition as $U_{n,m}=U_{n,m}^{(0)}\exp \left[ -i(c_{n}n+c_{m}m)/2%
\right] $, where $U_{n,m}^{(0)}$ is stationary configuration (\ref{omega}),
and vector $(c_{n},c_{m})$ determines the strength and direction of the
initial impulse. Figure \ref{fig3} demonstrates that the broad
small-amplitude soliton from Fig. \ref{fig2}(a), if kicked in the $n$
direction by the impulses of different strengths, $c_{n}=0.1$, $0.2$, and $%
0.5,$ is transformed into a tightly localized peak, which keeps nearly the
entire initial norm and comes to the halt at a site with coordinate $n_{f}$.
Note also that the initial broad soliton was unstable, while the final tall
peak (with almost the same norm) represents a stable soliton, as per the
black curve in Fig. \ref{fig1}.

\begin{figure}[th]
\epsfig{file=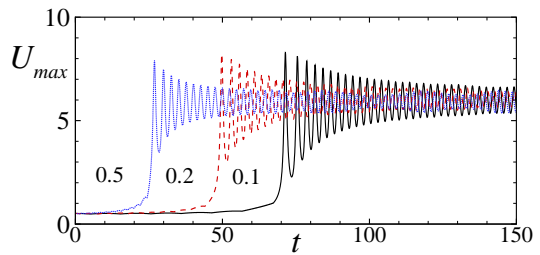,width=7cm}
\caption{(Color online) The evolution of the peak density for three
different values of the kick: $c_{n}=0.1$ (solid black), $c_{n}=0.2$ (dashed
red), and $c_{n}=0.5$ (dotted blue). Other parameters are as in Fig. \protect
\ref{fig3}.}
\label{fig3_max}
\end{figure}

This and other simulations corroborate the known fact that, under the action
of the kick applied in any direction relative to the underlying 2D lattice,
the broad 2D DNLS soliton starts to move, but eventually stops after having
traveled a finite distance (on the contrary to the 1D DNLS model, where the
kicked soliton may travel indefinitely long \cite{1D,Panos}). The harder the
initial shove $c_n$, the longer distance $n_f$ is passed by the discrete
soliton before the stoppage (cf. Fig.\ref{fig3}). However, an excessively
strong kick induces strong perturbations in the shape of the soliton, under
the action of which it starts to radiate, leading to a significant loss of
the norm. In Fig. \ref{fig3_max}, the peak density of the moving soliton,
determined as $U_{\max }=\max (|u_{n,m}|^{2})$, is displayed for different
strengths of the initial kick. It is seen that, after coming to the halt,
the initially broad soliton transforms into a highly localized mode whose
amplitude features gradually fading oscillations. Similar results were
obtained for the propagation of the discrete soliton initially kicked in the
diagonal direction, with $c_{n}=c_{m}$. 

Thus, persistent motion of solitons in the uniform 2D DNLS lattice
is impossible. This may be explained by the fact that, in the
continual limit, the cubic self-focusing leads to the collapse of
solitons; in the discrete setting, this implies the formation of
tightly localized tall peaks, which are strongly pinned to the
lattice, being therefore immobile. On the other hand, it is known
that motile 2D solitons are possible in discrete media with weaker
nonlinearities, which do not lead to the collapse in the continual
limit, \textit{viz}., saturable \cite{Johansson} and quadratic
\cite{chi2} on-site nonlinear terms. In the 2D lattice with the
combination of self-focusing cubic and defocusing quintic terms
(this combination of the nonlinearities does not give rise to the
collapse in the continuum medium either), the kicked soliton may
perform a long run, but eventually it comes to a halt, because of
radiation losses \cite{CQ}.

\subsection{Dragging the discrete solitons by the moving defect along the
linear trajectory}

\begin{figure}[th]
\epsfig{file=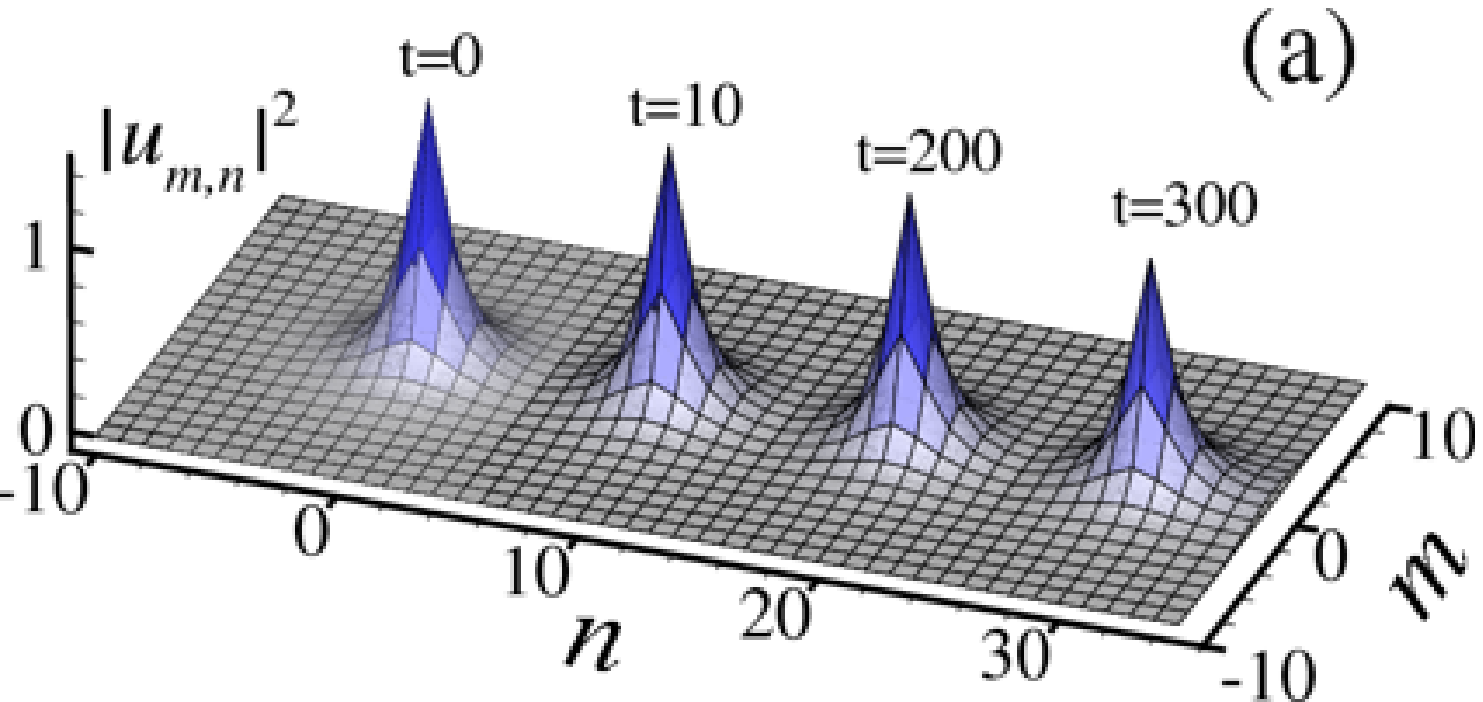,width=7cm}
\epsfig{file=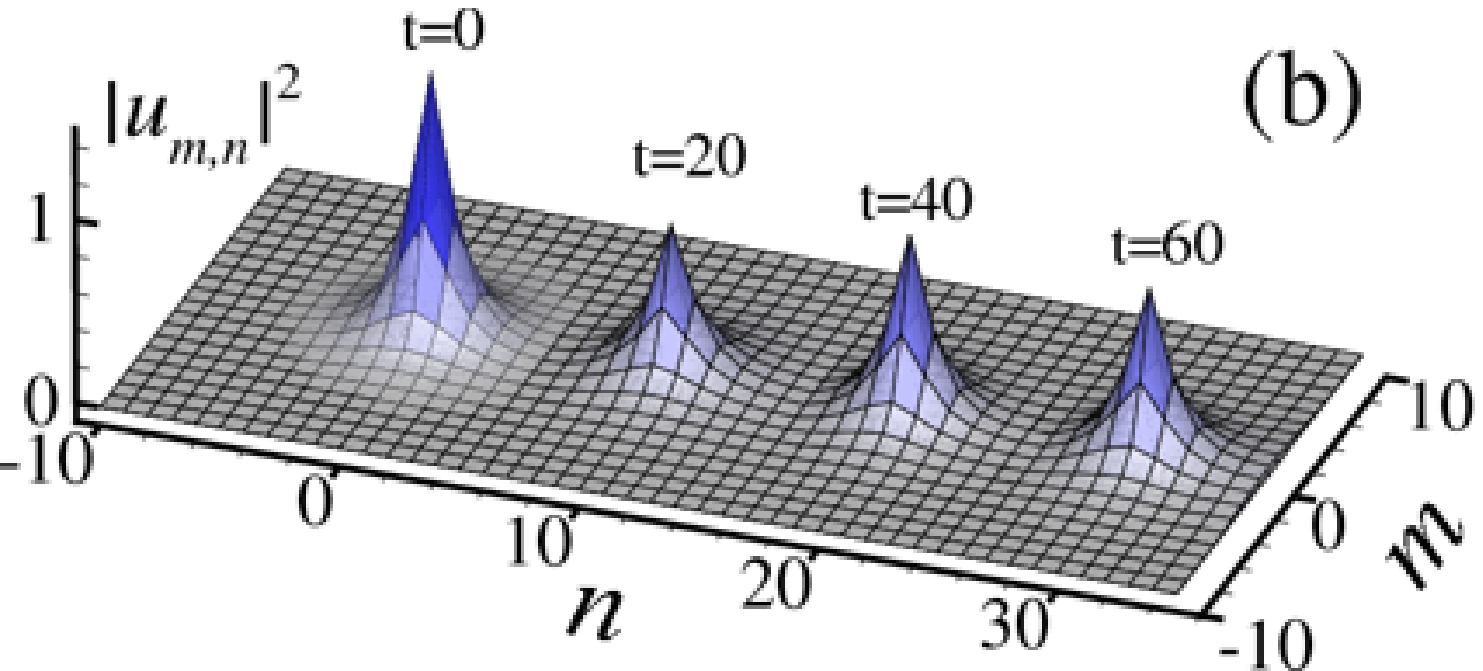,width=7cm}
\caption{Snapshots of the discrete soliton, dragged by the moving defect, at
different times. The initial profile, as well as parameters of the defect,
are the same as in Fig. \protect\ref{fig2}(b) ($v=\Delta =1$). The velocity
of the defect is $c_{n}=0.1$ in (a), and $c_{n}=0.5$ in (b). The other
component of the velocity is absent, $c_{m}=0$.}
\label{fig5}
\end{figure}

Forced motion of solitons attached to moving defects is the central topic of
this work. To demonstrate a typical realization of this scenario, we take
the initial soliton from Fig. \ref{fig2}(b), and drive the defect in the
lattice plane along a rectilinear rout: $n(t)=n_{s}+c_{n}t,$~$%
m(t)=m_{s}+c_{m}t$, where $c_{n}$ and $c_{m}$ are the velocities in the $n$
and $m$ directions. A set of typical examples is displayed in Fig. \ref{fig5}%
, for dragging the discrete soliton in the $n$ direction with different
velocities, $c_{n}=0.1,$ and $0.5$. This figure demonstrates that the
soliton can be transferred, virtually without any loss of the norm, over
indefinitely long distances, if the velocity, which is applied
instantaneously, is small enough, allowing the soliton to permanently adjust
itself, in an adiabatic manner, to the positions passed in the course of the
motion [Fig. \ref{fig5}(a)]. In more quantitative terms, this may be
explained by the comparison of the dragging velocity with the vectorial
group velocity produced by dispersion relation (\ref{disp}): $\mathbf{v}%
\equiv \partial \omega /\partial \left\{ k,q\right\} =2\left\{ \sin k,\sin
q\right\} $. If the dragging velocity is not small enough in comparison with
the group velocity, the adiabatic adjustment of the moving soliton is not
possible, and the soliton will be destroyed by the emission of radiation
waves escaping at the group velocity.

\begin{figure}[th]
\epsfig{file=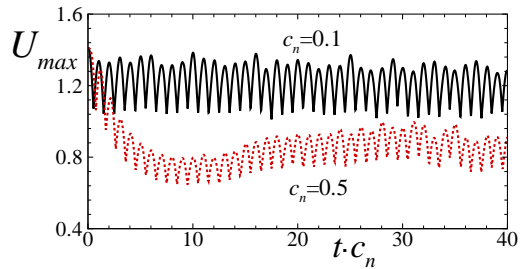,width=7cm}
\caption{(Color online) The evolution of the peak density for two different
values of the dragging velocity: $c_{n}=0.1$ (solid black), $c_{n}=0.5$
(dashed red). Other parameters are the same as in Fig.~\protect\ref{fig5}.}
\label{fig5_max}
\end{figure}

To further illustrate peculiarities of the dynamics of the driven soliton,
its peak density is displayed in Fig. \ref{fig5_max}, as a function of the
rescaled time, for different dragging velocities. The oscillations
correspond to transformations between on-site (maximum peak density) and
off-site (minimum peak density) configurations of the discrete soliton. At
small dragging velocities, the total norm remains practically constant after
a small initial loss of the norm due to the radiation, while the large
velocity ($c_{n}=0.5$ in Fig. \ref{fig5_max}) causes a significant initial
loss. In the latter case, the subsequent dynamics shows conservation of the
total norm. Similar results (not displayed here) have been obtained for the
driven motion of the soliton along the lattice diagonal. We do not aim here
to exactly identify the critical velocity for the transition from the
movable solitons to immovable ones, as the transition is somewhat fuzzy,
going through an intermediate region where the radiation loss suddenly
starts to increase with the growth of the driving velocity.

\subsection{Dragging the soliton along closed trajectories}

\begin{figure}[th]
\epsfig{file=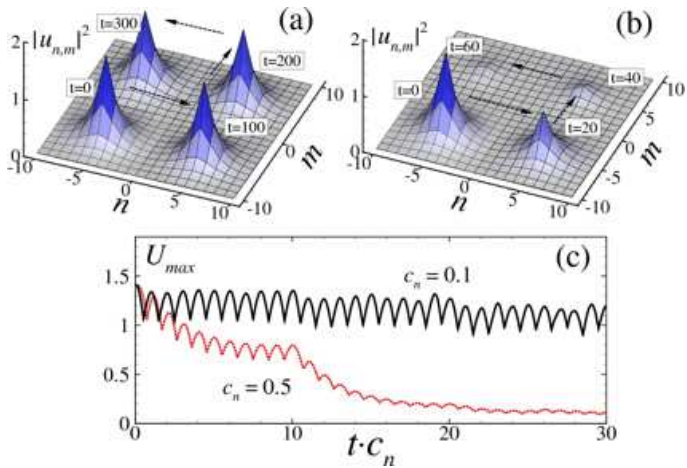,width=9cm}
\caption{The same as in Fig. \protect\ref{fig5}, but for dragging the
soliton along the square formed by the lattice axes. Here the corresponding
velocities along the $n$ and $m$ directions are $|c_{n}|=|c_{m}|=0.1$ in (a)
and $|c_{n}|=|c_{m}|=0.5$ in (b). In panel (c), the time evolution of the
peak density $U_{\max }$ for both cases is shown.}
\label{fig6}
\end{figure}

\begin{figure}[th]
\epsfig{file=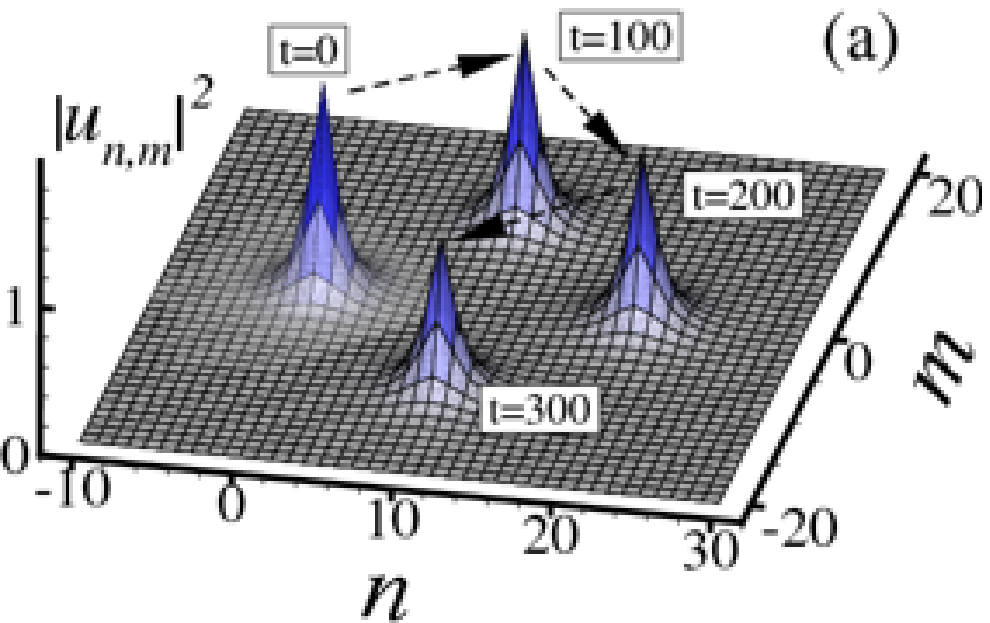,width=5cm} \epsfig{file=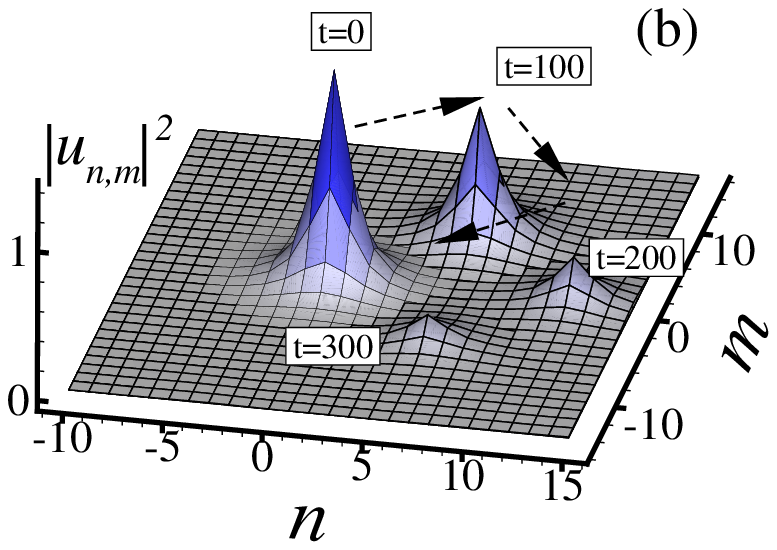,width=4.5cm}%
\epsfig{file=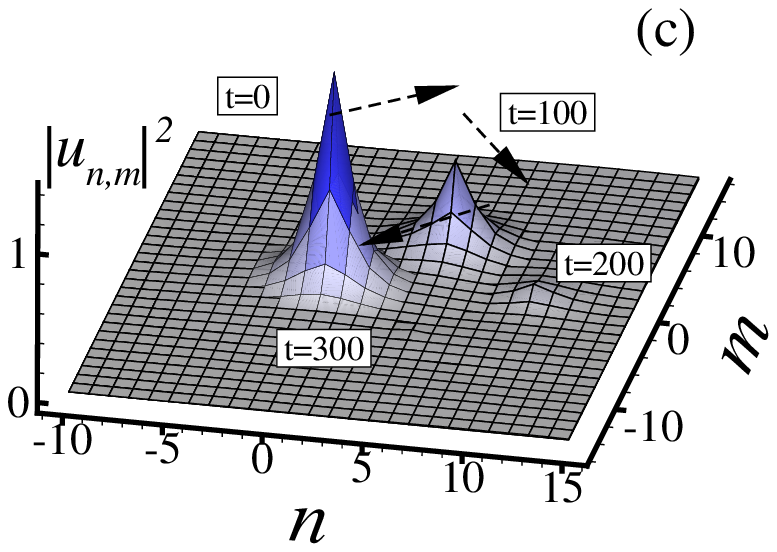,width=4.5cm} \caption{The same as in
Figs. \protect\ref{fig6}(a) and (b), for dragging
the soliton along the rhombus formed by diagonals of the lattice. Here $%
|c_{n}|=|c_{m}|=0.1$, $0.3$, and $0.5$, in (a), (b) and (c), respectively. }
\label{fig7}
\end{figure}

\begin{figure}[th]
\epsfig{file=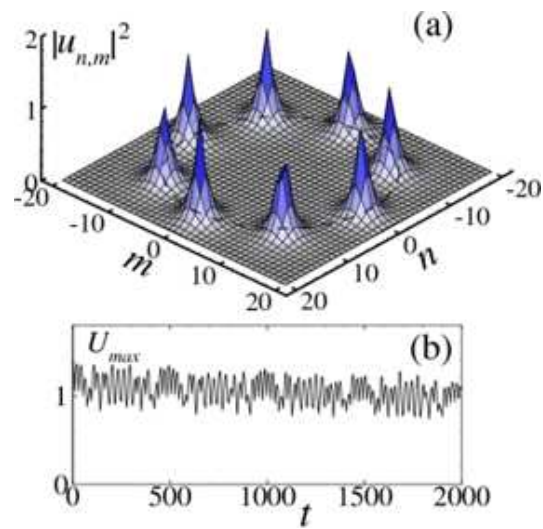,width=7cm}
\caption{Dynamics of the discrete soliton with initial profile taken from
Fig. \protect\ref{fig2}(a) and initially placed at position $(n,m)=(R,0)$
with $R=15$. The defect moves along the circular trajectory: $n_{d}=R\cos
(0.01t/\protect\pi )$ and $m_{d}=R\sin (0.01t/\protect\pi )$. The snapshots
are shown with interval $t=250$. In panel (b), the evolution of the peak
density is shown.}
\label{fig8}
\end{figure}

To demonstrate more sophisticated kinetic effects, in Fig. \ref{fig6} we
display dragging the soliton along a closed trajectory formed by four
segments aligned with bonds of the lattice. Typical results are also
displayed in Fig. \ref{fig7} for the driven motion along a closed route
formed by four diagonal segments. Figures \ref{fig6} and \ref{fig7} suggest
that, in the case of the abrupt changes in the direction of the driven
motion, the velocity of the dragging defect is a crucial parameter, leading
to strong losses at large velocities. Indeed, the comparison of Figs. \ref%
{fig6}(b) \ref{fig7}(c) to Fig. \ref{fig5}(b) demonstrates that relatively
high velocities produce a much more destructive effect on the solitons
dragged along the closed trajectories than on their linearly driven
counterparts. On the other hand, for sufficiently small velocities [$c_{n}=$
$0.1$ in Figs. \ref{fig6}(a) and \ref{fig7}(a)], the soliton can be dragged
along complex routes in the virtually intact form. Continuing the study in
this direction, we have also considered driven circular motion of the
soliton, as shown in Fig. \ref{fig8}, where the discrete soliton is safely
dragged along the ring trajectory with radius $R=15$ and angular velocity $%
0.01/\pi $.

\subsection{Collisions between discrete solitons dragged by defects}

\begin{widetext}
\begin{figure}[h]
\epsfig{file=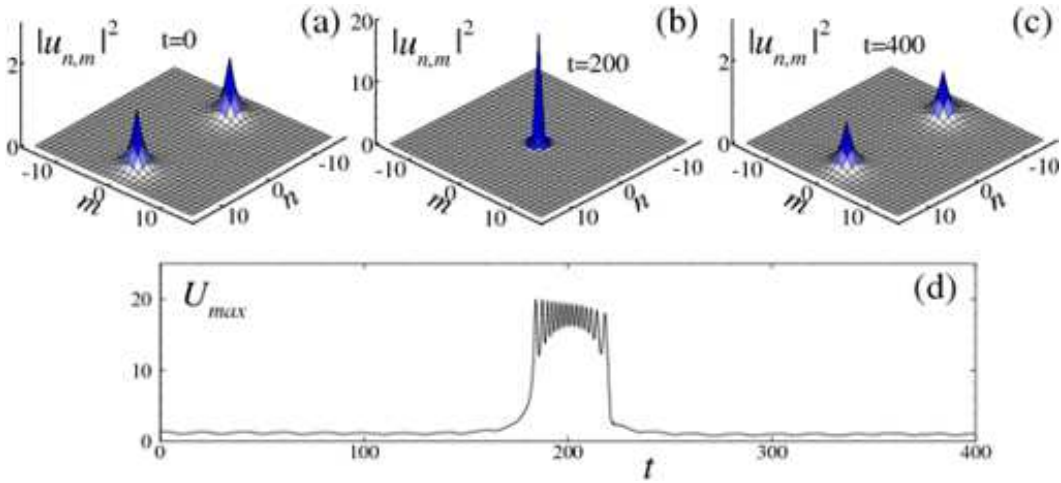,width=14cm}
\caption{The collision between two discrete solitons dragged by
defects with velocities $|c_{n}|=0.05$. Panels (a), (b) and (c)
display profiles of the solitons in the initial state, at the
collision moment, and in the final state, respectively. In panel (d)
the respective evolution of the peak density is presented. The
initial solitons are taken from Fig. \protect\ref{fig1}(b).}
\label{fig9}
\end{figure}
\end{widetext}

Another natural problem is collisions between defect-dragged
solitons. First, we consider the head-on collision between identical
solitons dragged toward each other. Colliding, they temporarily
merge into a ``lump" with the double norm, roughly adjusted to the
defect with doubled strength. Continuing the simulations and
monitoring attempts of the subsequent evolution of the merged lump,
we observed two different scenarios, depending on the velocity of
the dragging. If the velocity is higher then some critical value,
after the separation of the two defects the lump splits back into
two solitons, which are tantamount to those existing before the
collision, see Fig. \ref{fig9}.

\begin{widetext}
\begin{figure}[h]
\begin{center}
\epsfig{file=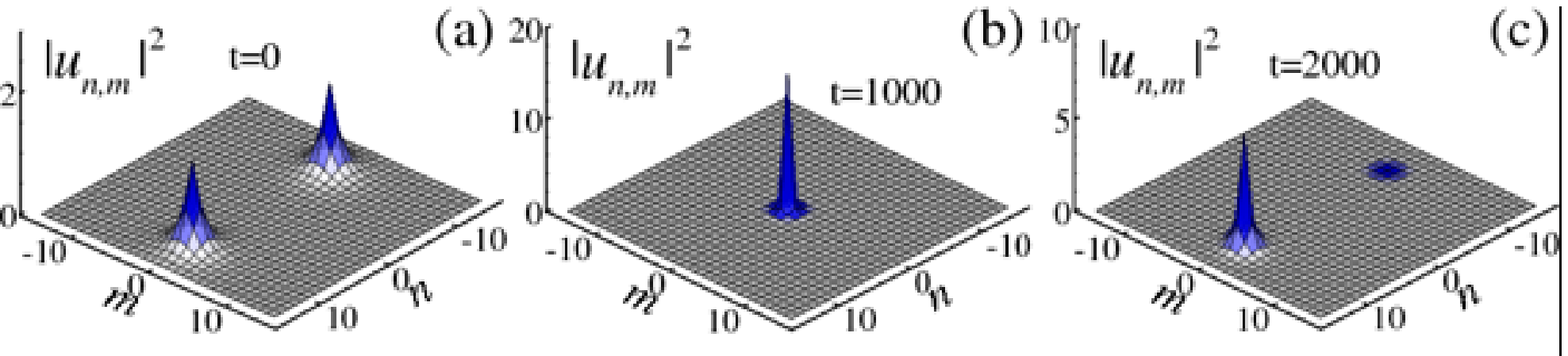,width=14cm}
\epsfig{file=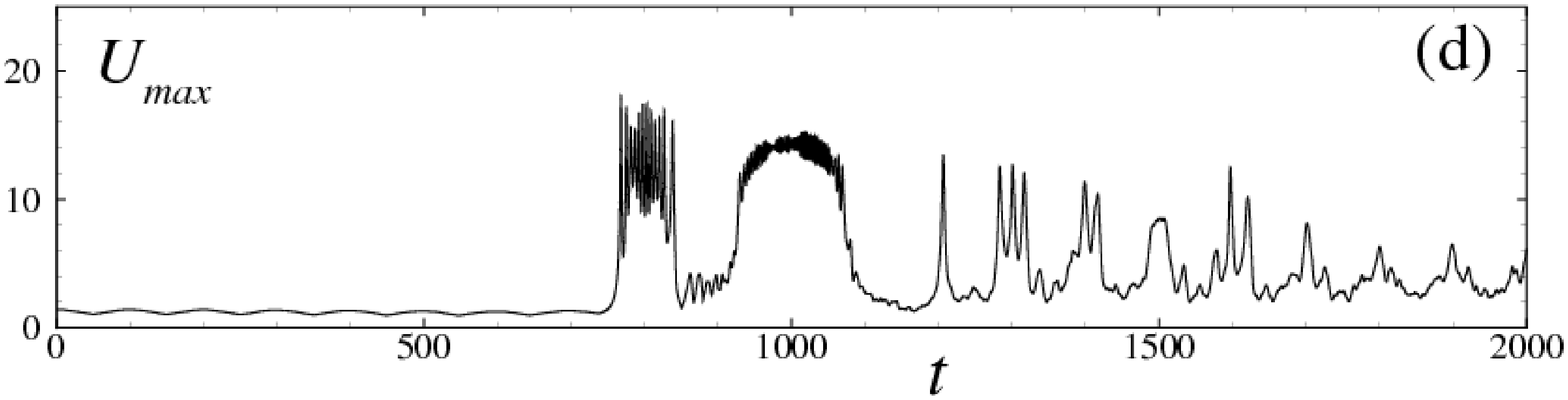,width=14cm}
\end{center}
\caption{The same as in Fig.\protect\ref{fig9}, but for velocities $%
|c_{n}|=0.01$.}
\label{fig10}
\end{figure}
\end{widetext}

The collision dynamics drastically changes if the velocity is
smaller than the critical value, see Fig. \ref{fig10}. Just before
the collision, but when the two defects are still well separated in
space (in the present case, at $t\approx 750$), one observes a
symmetry breaking in the distribution of the local power (density)
in the merging lump, which becomes unstable against oscillations of
the total norm between the two defects. At the time of the
collision, almost the same behavior occurs as in the previous
example pertaining to the higher velocity. However, after the
separation of the defects, the lump does not split, staying
attached, as a single soliton, to one defect and moving with it,
while the other defect remains ``bare", as seen in Fig.
\ref{fig10}(c).

\begin{figure}[h]
\epsfig{file=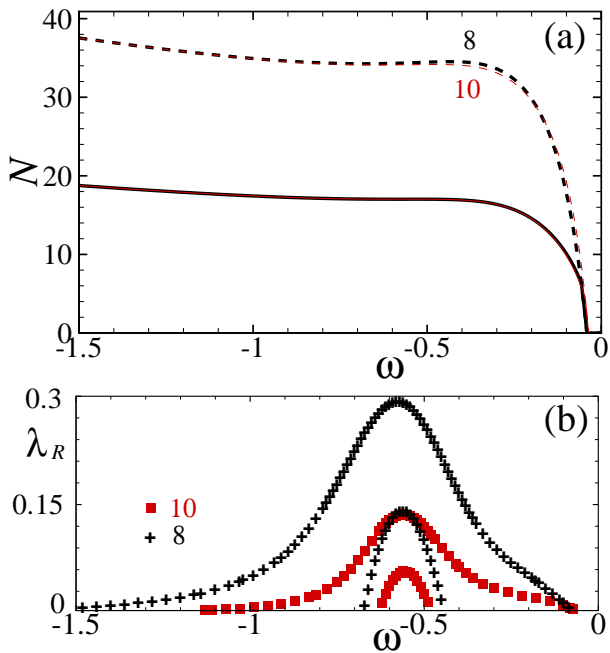,width=8cm}
\caption{(Color online) (a) Existence curves for the families of symmetric
(upper lines) and asymmetric (lower lines) discrete solitons supported by a
symmetric pair of the defects. Solid and dashed portions of the curves
correspond to stable and unstable solutions. Black and red lines (mostly
overlapping) pertain to the configurations with the distance between the
defects fixed to be $\protect\delta =8$ and $10$ sites, respectively. (b)
The frequency dependence of the real part of the two most unstable eigenvalues
for the symmetric solutions from (a). The black (\textbf{+}) and red ($%
\blacksquare $) symbols correspond to the distances between the defects $%
\protect\delta =8$ and $10$, respectively. }
\label{stab}
\end{figure}

The symmetry breaking in the head-on collisions of the dragged solitons can
be explained by the consideration of a linear-stability problem. To this
end, we calculated the existence curves for static symmetric and asymmetric
modes trapped by a pair of defects separated by different \emph{constant}
distances, and checked the linear stability of those modes, taking perturbed
solutions as $u_{n,m}=\exp (-i\omega t)[U_{n,m}+a_{n,m}\exp (\lambda
t)+b_{n,m}\exp (\lambda ^{\ast }t)]$, where $\lambda $ is an eigenvalue, $%
U_{n,m}$ is the unperturbed stationary solution, and $a_{n,m},b_{n,m}$ are
small perturbations. The so found unstable symmetric (dashed lines) and
stable asymmetric (solid lines) families of the solutions are presented in
Fig. \ref{stab}(a), with the distance between the defects being $\delta =8$
(black) and $10$ (red) sites (the corresponding existence curves are
practically overlapping). In Fig. \ref{stab}(b) dependences of the most
unstable eigenvalues (the real part of $\lambda $) on frequency $\omega $
for the symmetric configurations with different distances between the
defects from Fig. \ref{stab}(a) are presented. One can conclude that the
symmetric configuration is unstable against small perturbations (with the
most unstable eigenvalue substantially diminishing with the increase of the
separation between the defects). Although this instability pertains to the
symmetric lumps supported by the pair of immobile defects, it provides an
explanation to the instability of the merged symmetric lump also in the case
of the slowly colliding defects.

As mentioned above, the symmetry breaking occurs between the colliding
solitons driven by the slowly moving defects. This fact can be explained by
comparing the time necessary for the development of the instability of the
lump produced by the collision and the collision time. The time of the
mutual passage of the solitons attached to rapidly moving defects is smaller
than the time required for the onset of the instability. In this case, the
collision is quasi-elastic, and it does not cause the breaking of the
symmetry between the solitons, while for smaller velocities the collision
time is sufficient for the development of the instability and ensuing
symmetry breaking, which may eventually transform the two solitons into one,
pinned to a single defect. Similar arguments were previously used to explain
the transition from quasi-elastic collisions to symmetry-breaking ones in
several models supporting 1D solitons in continual models \cite{Javid}.

We have also considered the collision of solitons with a mismatch in
direction $m$ perpendicular to the velocities (the ``aiming
parameter"), $\xi =|m_{1}-m_{2}|$, where $m_{j}$ defines the initial
coordinate of the $j$-th soliton in the direction of $m$. In
particular, for the same small velocities as in Fig. \ref{fig10},
with two different values
of the mismatch, $\xi =8$ and $\xi =10$, the results are shown in Fig. \ref%
{fig11}. It is seen that the character of the collision can be controlled by
varying the mismatch, cf. Ref. \cite{George}, where the collisions were
simulated between 2D solitons in the framework of a continual dissipative
equation.

\section{Conclusion}

In this work, we have elaborated a scenario for the controlled transfer of
2D discrete solitons through the DNLS lattice. The main obstacle to the
transport is the fact that 2D solitons in the uniform DNLS system cannot
move persistently under the action of an initial kick, being always braked
by the underlying Peierls-Nabarro potential. Nevertheless, we have
demonstrated that a relatively broad soliton can be dragged over
indefinitely large distances, with virtually zero loss, by the moving
attractive defect, provided that the dragging velocity does not exceed a
critical value. This is qualitatively explained by the comparison of the
dragging velocity with the group velocity of the linear waves propagating in
the uniform lattice. The stable driven motion is possible in any direction,
as well as along complex routes with corners, and along circular
trajectories; however, the critical velocity is lower in the latter cases.
Collisions between solitons dragged by two solitons in opposite directions
were considered too, with the conclusion that the two solitons spontaneously
merge into a single one, which stays attached to either moving defect, if
the collision velocity is small enough.

These scenarios can be implemented (and used for various applications) in
the following form: a quiescent soliton may be prepared in the uniform 2D
lattice, then the local attractive defect(s) may be induced [for instance,
by laser beam(s) illuminating the corresponding BEC], effectively converting
the free solitons into defect modes; next, the mode(s) may be transferred to
a new position, as described above, and, eventually, the laser beam(s) may
be switched off. Eventually, free solitons may be transferred according to
this protocol, in the medium where these solitons are not motile by
themselves.

The analysis presented above may be naturally extended in various
directions, including the transfer of vortex solitons \cite{vortex} (in the
latter case, the trapping defect must be wide enough, for a sufficient
overlap with the vortex; in fact, the defect itself may have a vortical
structure). A challenging generalization would be to develop a similar
scenario for discrete solitons in 3D lattices, which may also be realized in
terms of BEC.

\begin{widetext}
\begin{figure}[th]
\begin{center}
\epsfig{file=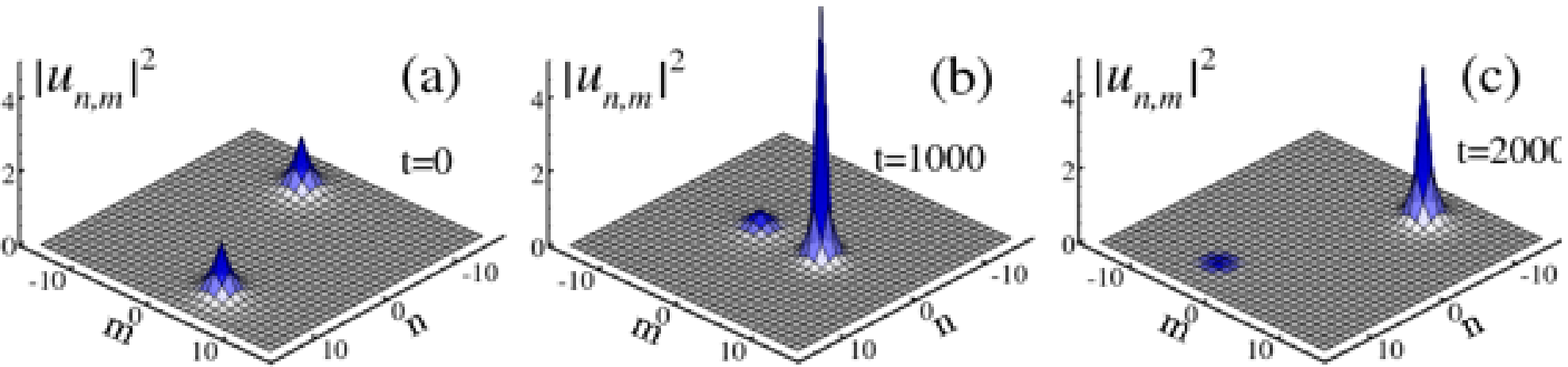,width=14cm}
\epsfig{file=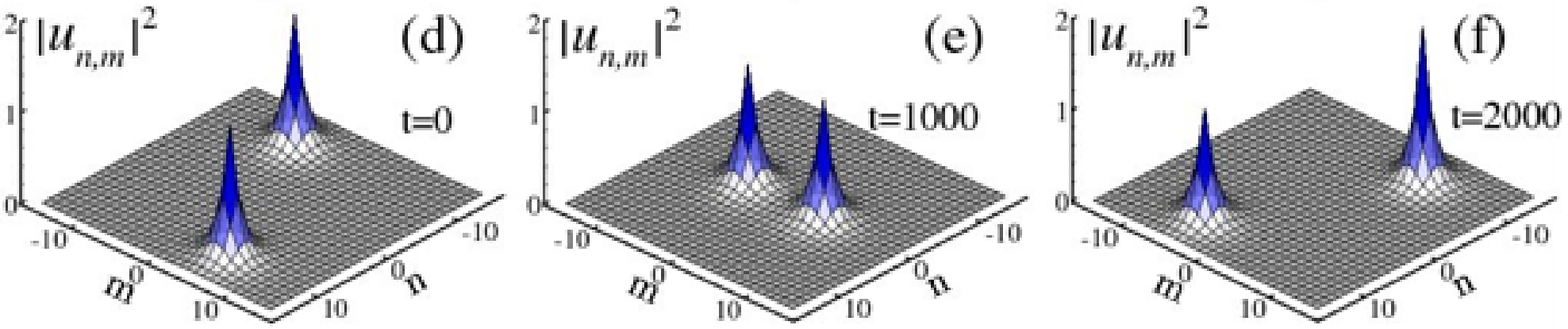,width=14cm}
\epsfig{file=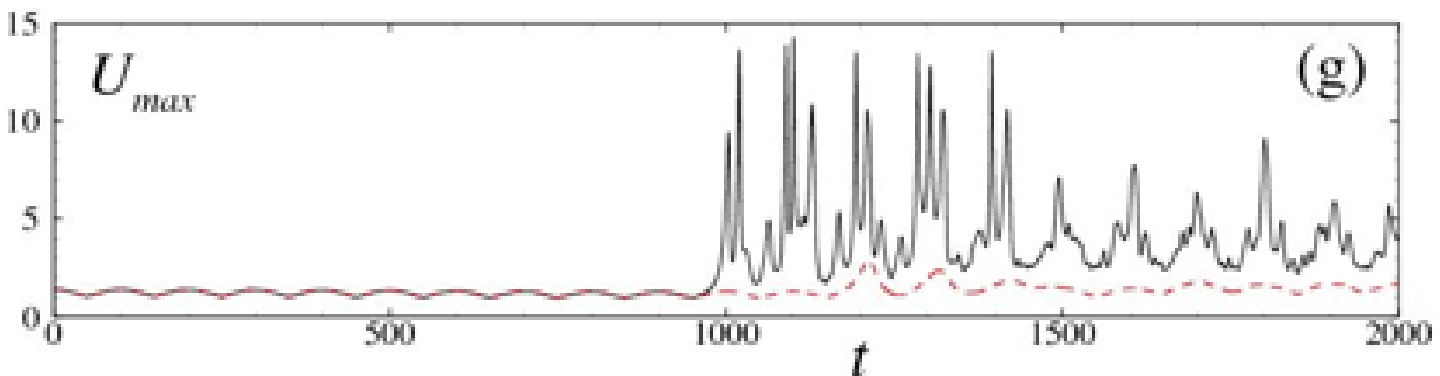,width=14cm}
\end{center}
\caption{Color online) The collision with finite mismatch
$\protect\xi $. The snapshots in (a)-(c) and (d)-(f) correspond to
$\xi =8$ and $\xi=10$, respectively. In panel (g), the respective
evolution of the peak density is shown for $\xi =8$ (solid black)
and $\xi =10$ (dashed red).} \label{fig11}
\end{figure}
\end{widetext}

\section*{Acknowledgments}

V.A.B. acknowledges the support from the FCT grant, PTDC/FIS/64647/2006.
B.A.M. appreciates hospitality of Centro de F\'{\i}sica do Porto (Porto,
Portugal).

\end{document}